\documentstyle[12pt]{article}

\begin{document}

\setcounter{page}{1}

\begin{titlepage}
\title{
\begin{flushright}
{\bf\normalsize COLO-HEP-279}\\
\end{flushright}
ISING MODEL COUPLED TO THREE-DIMENSIONAL QUANTUM GRAVITY\\
}
\author{
C.F. Baillie\\
Physics Department\\
University of Colorado\\
Boulder, CO 80309\\
}
\date{\today}
\end{titlepage}

\maketitle

\begin{abstract}
We have performed Monte Carlo simulations of the Ising model
coupled to three-dimensional quantum gravity
based on a summation over dynamical triangulations.
These were done both
in the microcanonical ensemble, with the number of points in
the triangulation and the number of Ising spins fixed, and in the grand
canonical ensemble.
We have investigated the two
possible cases of the spins living on the vertices of the triangulation
(``direct'' case)
and the spins living in the middle of the tetrahedra (``dual'' case).
We observed phase transitions which are probably second order,
and found that the dual implementation more effectively couples the spins to
the
quantum gravity.
\end{abstract}
\eject

\section{Introduction}

Following recent Monte Carlo simulations of spin models coupled to
two-dimensional
quantum gravity \cite{spin2d} and of pure three-dimensional quantum gravity
\cite{pure3d}
we have performed some small exploratory
simulations of the Ising model coupled to three-dimensional quantum gravity.
This is the simplest model of matter coupled to quantum gravity in three
dimensions, and provides
a step on the way to simulating the physical case of matter coupled to
four-dimensional quantum gravity.

The continuum Einstein action for 3-dimensional gravity is
\begin{equation}
  S_{E}= \int d^3\xi\sqrt{g}
        \left( \lambda - \frac{R}{2\pi G} \right),
\label{EG}
\end{equation}
where $\lambda$ is the cosmological constant and $G$ is Newton's constant.
When a dynamical triangulation (with $N_0$ vertices, $N_1$ links, $N_2$
triangles
and $N_3$ tetrahedra) is used to regularize the functional integration
over metrics the two terms in (\ref{EG}) discretize as follows:
\begin{equation}
  \int d^3\xi\sqrt{g} \rightarrow N_3
\end{equation}
\begin{equation}
  \int d^3\xi\sqrt{g}R \rightarrow \sum_l \left( c - n(l) \right),
\label{NL}
\end{equation}
where $n(l)$ is the number of tetrahedra which share the link $l$,
and $c=2\pi/\alpha$, with $\alpha=arccos({1 \over 3})$,
is a number which ensures that $R$ is zero for flat space.
Since each tetrahedron is comprised of six links, $\sum_l n(l) = 6 N_3$
and (\ref{NL}) can be written as $c N_1 - 6 N_3$.
Hence the discrete version of (\ref{EG}) is
\begin{equation}
  S = \lambda N_3 - \frac{1}{2\pi G} \left( c N_1 - 6 N_3 \right).
\label{SD}
\end{equation}
Now since we have the following relations for a three-dimensional
triangulation (or ``three-dimensional simplicial manifold'')
\begin{equation}
  N_3 - N_2 + N_1 - N_0 = 0
\label{E}
\end{equation}
\begin{equation}
  N_2 = 2 N_3
\label{T}
\end{equation}
(first is Euler's relation; second is because each triangle is common to
two tetrahedra), we can change to the more convenient variables $N_0$, $N_3$
and write
\begin{equation}
  S = \lambda_3 N_3 - \lambda_0 N_0,
\label{SF}
\end{equation}
with $\lambda_3 = (\lambda + \frac{6-c}{2\pi G})$
and $\lambda_0 = \frac{c}{2\pi G}$.
Therefore the grand canonical partition function for quantum gravity in three
dimensions based on a summation over dynamical triangulations $T$
with $N_3$ tetrahedra and $N_0$ vertices is
\begin{equation}
  Z(\lambda_3,\lambda_0) =
  \int dN_3 dN_0 Z_{N_3,N_0} e^{-\lambda_3 N_3 + \lambda_0 N_0},
\label{ZGC}
\end{equation}
where
\begin{equation}
  Z_{N_3,N_0} = \sum_{T(N_3,N_0)} \rho(T)
\label{ZGC}
\end{equation}
is the micro-canonical partition function in which $N_3,N_0$ are fixed,
and $\rho(T)$ is a suitable weight for each triangulation $T$.
One usually appeals to universality and sets $\rho(T)=1$.

In order to perform a Monte Carlo simulation of either of
these partition functions for pure three-dimensional quantum
gravity one must define a set of ``moves'' which ergodically sum over all
possible triangulations. In two dimensions this is straight-forward since
one move suffices for each case: for the grand canonical partition function
one uses the so-called ``split-join'' move, and for the micro-canonical
partition
function the ``flip''. Unfortunately things are not as simple in three
dimensions --
there is no known set of moves which is ergodic in the micro-canonical
ensemble;
however two moves exist for the grand canonical case: the so-called ``Alexander
move''
which is a generalization of the two-dimensional split-join move changing one
tetrahedron
into four tetrahedra (or vice versa), and a generalization of the `flip''
taking two tetrahedra
into three tetrahedra (or vice versa).
For more details and illustrations of these moves see \cite{pure3d}.

If we now couple the Ising model to this discretization of three-dimensional
quantum gravity the action (\ref{SF}) becomes
\begin{equation}
  S_I = \lambda_3 N_3 - \lambda_0 N_0 + \beta \sum_{i,j}S_{ij}(1 -
\sigma_i\sigma_j),
\label{SI}
\end{equation}
where $S_{ij}$ is the connectivity matrix of the triangulation, $\sigma_i$ are
the Ising spins
taking values $+1$ or $-1$, and $\beta$ is the inverse temperature.
There are two possible ways in which to add the Ising spins to the dynamical
triangulation: they can either live on the vertices in which case there will be
$N_0$ spins or they can live in the tetrahedra giving $N_3$ spins. We shall
refer to the former as the ``direct'' case and the latter as the ``dual''.
For the direct implementation there are $N_1 = N_0+N_3$ links between the
spins so $S_{ij}$ will contain this number of unit entries; for the dual case
there
are $N_2 = 2 N_3$ interactions, and hence non-zeros in $S_{ij}$.
Note that we have arranged the normalization such that
at zero temperature, i.e. $\beta = \infty$, all the Ising spins line up
and their contribution to $S_I$ vanishes.
At $\beta=0$ we recover pure quantum gravity.

As for the pure three-dimensional quantum gravity case, there are two possible
partition functions: grand canonical and micro-canonical. For the grand
canonical
ensemble, one must create and destroy Ising spins when either vertices in the
direct case
or tetrahedra in the dual case are created and destroyed. Destruction of Ising
spins is
trivial -- they are simply removed -- however creation is rather more subtle --
the values of the new spins should be picked randomly to avoid any bias.
Our simulation of the grand canonical Ising plus three-dimensional quantum
gravity
uses the same methods as were used in all the pure three-dimensional quantum
gravity
simulations, see \cite{pure3d} for details.

As a step on the way to the grand canonical simulation, we have chosen to
simulate the
micro-canonical ensemble. To do this $N_3,N_0$ and the number of Ising spins
must be fixed.
Therefore we use only the flip move to update the triangulation, and we use it
in pair-wise fashion --
firstly changing two tetrahedra to three tetrahedra then immediately after
(somewhere else in
the triangulation) changing three tetrahedra to two tetrahedra. As explained
above this is
not known to be ergodic for the pure three-dimensional quantum gravity case,
so we cannot assume it to be when we couple in Ising spins. However it is an
easier simulation to do and may provide an approximation to the more realistic
grand canonical case (this is somewhat like simulating quenched QCD as an
approximation to full QCD in lattice gauge theory).

For both the grand canonical and micro-canonical simulations, in addition to
updating
the triangulation, we must update the Ising spins.
To do this we make use of cluster algorithms, introduced
in \cite{SW}, which are very effective
in the vicinity of phase transitions, where
spins tend to form large clusters. However, away from the
transition, the standard  Metropolis method is more practical.
Therefore, we used both algorithms to update the Ising spins, doing one pass of
each per update step.
We actually used Wolff's version of the cluster
algorithm \cite{wolf}, as it is faster than Swendsen-Wang's \cite{SW}.

\section{Micro-canonical}

Physically the vacuum in four-dimensional gravity is almost flat, therefore
in our micro-canonical simulations of three-dimensional quantum gravity we
pick $N_3,N_0$ so that the average curvature of the triangulation
\begin{equation}
  \bar R = {2\pi (N_3+N_0) - 6\alpha N_3 \over N_3}
\label{Ra}
\end{equation}
is as small as possible; which in turn implies that ${N_0 \over N_3} \approx
.1755$.
This leads to the following values of $N_0,N_3$: 69,393 ; 154,878 and 325,1852.
If $n_3^i$ is the number of tetrahedra meeting at the vertex $i$ then, since
each
tetrahedron has four vertices, $\sum_i n_3^i = 4 N_3$, and therefore
$<n_3> = 4 {N_3 \over N_0} \approx 22.8$. Thus each vertex is shared by roughly
$23$
tetrahedra. This means that for the direct implementation of the Ising spins on
the
triangulation, each spin will have on average $23$ neighbors, whereas for the
dual
implementation each spin will have only four neighbors (the number of faces of
a tetrahedron).
Therefore we can expect a much stronger correlation of spins for a given
$\beta$ for the direct
case compared to the dual case, which causes the significantly stronger
phase transition to occur at a lower critical value of $\beta$.
We shall first present our results for the direct case and then for the dual
case.
In both cases we measured the standard thermodynamic energy $E$ and
magnetization $M$ for the Ising model, the acceptance rate
of the Metropolis algorithm $a_M$, the acceptance rate for the flip moves
$a_f$, and
the fractal dimension of the spin clusters constructed by the Wolff algorithm.

For the direct case the number of Ising spins is $N_0 = 69, 154$ and $325$.
We ran simulations at the following 18 values of $\beta$:
$0.01, 0.02, 0.03, 0.0325, 0.035, 0.0375, 0.04, 0.0425, 0.045, 0.0475,$
$0.05, 0.06, 0.07, 0.08, 0.09, 0.1, 0.2, 0.3$. The results for $E,M,a_M,a_f$
for the
$N_0=325$ simulation are listed in Table 1.

\begin{center}
\begin{tabular}{|c|c|c|c|c|} \hline
$\beta$ & $E$     & $M$      & $a_M$    & $a_f$    \\[.05in]
\hline
 0.01   & 0.99(6) & 0.047(1) & 0.973(1) & 0.995(1) \\[.05in]
 0.02   & 0.98(3) & 0.054(1) & 0.943(1) & 0.990(1) \\[.05in]
 0.03   & 0.95(2) & 0.077(1) & 0.904(1) & 0.985(1) \\[.05in]
 0.0325 & 0.94(2) & 0.083(1) & 0.890(1) & 0.984(1) \\[.05in]
 0.0350 & 0.92(2) & 0.098(2) & 0.874(1) & 0.983(1) \\[.05in]
 0.0375 & 0.90(2) & 0.124(2) & 0.852(1) & 0.982(1) \\[.05in]
 0.04   & 0.86(3) & 0.154(5) & 0.824(1) & 0.980(1) \\[.05in]
 0.0425 & 0.82(1) & 0.187(2) & 0.788(1) & 0.980(1) \\[.05in]
 0.0450 & 0.75(1) & 0.236(1) & 0.751(1) & 0.979(1) \\[.05in]
 0.0475 & 0.71(1) & 0.266(1) & 0.715(1) & 0.978(1) \\[.05in]
 0.05   & 0.66(1) & 0.301(2) & 0.683(1) & 0.978(1) \\[.05in]
 0.06   & 0.53(1) & 0.384(1) & 0.592(1) & 0.976(1) \\[.05in]
 0.07   & 0.45(1) & 0.444(2) & 0.529(1) & 0.975(1) \\[.05in]
 0.08   & 0.40(1) & 0.487(1) & 0.480(1) & 0.974(1) \\[.05in]
 0.09   & 0.35(1) & 0.529(1) & 0.440(1) & 0.973(1) \\[.05in]
 0.10   & 0.32(1) & 0.566(1) & 0.403(1) & 0.973(1) \\[.05in]
 0.20   & 0.11(1) & 0.840(1) & 0.160(1) & 0.980(1) \\[.05in]
 0.30   & 0.04(1) & 0.930(1) & 0.067(1) & 0.988(1) \\[.05in]
\hline
\end{tabular}
\end{center}
{\narrower\smallskip\noindent
Table 1: Measured values of energy $E$, magnetization $M$,
Metropolis acceptance $a_M$ and flip acceptance $a_f$ from $N_0=325$ direct
simulation.
\smallskip}
\bigskip

\noindent
By numerically differentiating the energy and magnetization we obtain the
specific heat
and susceptibility shown for all $N_0$ in Figs. 1 and 2 respectively.
Using a cubic spline approximation we estimate the positions $C_{max}$ and
$\chi_{max}$ of the maximum of the peaks to
obtain the critical inverse temperatures $\beta_c$ for each $N_0$ listed in
Table 2.

\begin{center}
 \begin{tabular}{|c|c|c|} \hline
  $N_0$  & $\beta_c$ for $C_{max}$ & $\beta_c$ for $\chi_{max}$ \\ \hline
  69  & 0.071(1) & 0.069(1) \\
  154 & 0.056(1) & 0.055(1) \\
  325 & 0.050(2) & 0.046(2) \\
  \hline
 \end{tabular}
\end{center}
{\narrower\smallskip\noindent
Table 2: Position of peaks in specific heat and susceptibility
for each value of $N_0$ from direct simulation.
\smallskip}
\bigskip

\noindent
We could attempt to fit to the specific heat and susceptibility peaks in order
to
extract exponents and/or we could try finite-size scaling analysis but as these
are
exploratory simulations of small systems the results would not be very
impressive
or reliable.
Therefore we leave this for future calculations. Here we can say only that
there is
a phase transition, at approximately $\beta_c=0.048(2)$ for $N_0=325$,
which is probably second order since the peaks in $C$ and $\chi$
increase with system size (first order is ruled out as there are no
discontinuites in $E$ and $M$).
Unfortunately without the critical exponents it is not possible
to extrapolate the $\beta_c$ values to $N_0 = \infty$.

Turning to the acceptance rates we see that the Metropolis acceptance falls off
as $\beta$
increases as expected -- the spins become frozen as the temperature falls.
However the flip acceptance rate does not change very much from its initial
value of $1$
at $\beta = 0$ (where the spins are random so do not systematically effect the
triangulation).
In other words the Ising spins do not seem to interact very strongly with
three-dimensional
quantum gravity when they are implemented directly -- i.e. live on the vertices
of the triangulation.
Numerically this is obviously due to the fact that during the flip move only
one out of on average $23$ Ising spins at each vertex is reconnected so
the energy change is relatively small.

Now we turn to the results for the dual case, where the number of Ising spins
is
$N_3 = 393,878$ and $1852$.
Simulations were done at 12 values of $\beta$:
$0.1,0.2,0.3,0.325,0.35,0.375,$
$0.4,0.425,0.45,0.475,0.5,0.6$;
the results from the $N_3=878$ simulation are in Table 3.

\begin{center}
\begin{tabular}{|c|c|c|c|c|} \hline
$\beta$ & $E$     & $M$      & $a_M$    & $a_f$    \\[.05in]
\hline
 0.1   & 0.894(7) & 0.033(1) & 0.846(1) & 0.782(1) \\[.05in]
 0.2   & 0.774(3) & 0.044(1) & 0.682(1) & 0.569(1) \\[.05in]
 0.3   & 0.639(2) & 0.068(2) & 0.513(1) & 0.376(1) \\[.05in]
 0.325 & 0.599(2) & 0.084(2) & 0.470(8) & 0.333(7) \\[.05in]
 0.350 & 0.560(1) & 0.105(2) & 0.427(9) & 0.291(7) \\[.05in]
 0.375 & 0.516(1) & 0.148(3) & 0.381(9) & 0.250(8) \\[.05in]
 0.4   & 0.456(1) & 0.254(5) & 0.327(1) & 0.206(1) \\[.05in]
 0.425 & 0.352(1) & 0.539(5) & 0.240(7) & 0.147(9) \\[.05in]
 0.450 & 0.255(1) & 0.728(2) & 0.168(9) & 0.106(5) \\[.05in]
 0.475 & 0.191(1) & 0.818(2) & 0.121(8) & 0.083(3) \\[.05in]
 0.5   & 0.146(1) & 0.872(1) & 0.088(1) & 0.068(1) \\[.05in]
 0.6   & 0.052(1) & 0.960(1) & 0.029(3) & 0.043(1) \\[.05in]
\hline
\end{tabular}
\end{center}
{\narrower\smallskip\noindent
Table 3: Measured values of energy $E$, magnetization $M$,
Metropolis acceptance $a_M$ and flip acceptance $a_f$ from $N_3=878$ dual
simulation.
\smallskip}
\bigskip

\noindent
We show the specific heat and susceptibility,
obtained from numerical differentiation, for $N_3=393,878$ (we do not have
enough
data for $N_3=1852$), in Figs. 3 and 4 respectively.
And we estimate the position of the maximum of the peaks -- Table 4.

\begin{center}
 \begin{tabular}{|c|c|c|} \hline
  $N_3$  & $\beta_c$ for $C_{max}$ & $\beta_c$ for $\chi_{max}$ \\ \hline
  393  & 0.444(1) & 0.445(1) \\
  878  & 0.421(2) & 0.416(2) \\
  \hline
 \end{tabular}
\end{center}
{\narrower\smallskip\noindent
Table 4: Position of peaks in specific heat and susceptibility
for two values of $N_3$ from dual simulation.
\smallskip}
\bigskip

\noindent
Again it is not worth attempting to fit to this data to extract critical
exponents.
For now we can say that we have seen a second order phase transition at
approximately $\beta_c=0.418(2)$ for $N_3=878$.

Thus for the dual case, the Ising spins significantly affect the
triangulation - the acceptance rate for the flip move drops from near $1$ at
low $\beta$
to almost $0$ as $\beta$ increases above $\beta_c$.
Thus matter in the form of Ising spins interacts strongly with quantum gravity
when it
is coupled to it through the dual of the triangulation -- i.e. when the spins
live in the
tetrahedra.
If we compare the direct and dual cases we see that for the former the
peaks in $C$ and $\chi$ are much larger as expected. Interestingly the
critical value of $\beta$ for the dual case in which each spin is
connected to four neighbors ($\beta_c=0.418(2)$) is closer to the value
($0.221652(3)$ \cite{ising3d}) for the standard 3D Ising model which has
coordination number six, than is the value ($0.048(2)$) from the direct case
with
its average coordination number of $23$.

Finally, we look at the fractal dimension of the clusters constructed by the
Wolff algorithm.
This is interesting because, as discussed by Meo, Heermann and Binder
\cite{MHB}, these
clusters are actually the real, physical clusters in the
system. For a standard Ising model simulated in $D$ dimensions, the fractal
dimension $d$ is
given by $d = D - \beta / \nu$ which is approximately $2.5$ for $D=3$
\cite{Stauff}.
In the simulations we stored the sizes and diameters~\footnote{By diameter
we mean the distance from the random starting point
used in construction of the cluster to the most remote point on the cluster
border;
on average this should be proportional to the gyration radius.}
of the Ising clusters.
The $\log-\log$ plots of the average cluster size versus it's
diameter along with the best fits for $d$, for both the direct and dual
simulations,
are shown in Fig. 5.
For the direct case we get a value of $d=2.7(2)$ close to the standard 3D Ising
model value,
for the dual case it is a little smaller ($2.0(1)$). These values should be
considered as preliminary
given the small triangulations used - the maximum sized clusters which fit on
the direct and
dual triangulations are only about 50 and 100 sites respectively.

\section{Grand canonical}

Armed with the knowledge that the dual implementation of the Ising spins on the
triangulation
more effectively couples them to three-dimensional quantum gravity in the
micro-canonical ensemble,
we performed a modest simulation of the dual case in the grand canonical
ensemble.
To do this one first picks likely values of the parameters $\lambda_3,
\lambda_0$ plus
widths $\Delta\lambda_3, \Delta\lambda_0$ then lets the Monte Carlo simulation
vary the former
within the latter and find the actual values of $\lambda_3, \lambda_0$.
We started with $\lambda_3 = 1, \lambda_0 = 0,
\Delta\lambda_3 = \Delta\lambda_0 = 0.25$
and ran the first simulation with $N_3 = 393$
at $\beta = 0.1, 0.2, 0.3, 0.35, 0.4, 0.45, 0.5$.
This resulted in the predictions for $\lambda_3, \lambda_0$ listed in columns 2
and 3
respectively in Table 5. Then we used these predictions as start values for the
second
simulation with $N_3 = 878$ resulting in the further predictions listed in
columns 4 and 5.
Also listed in Table 5, in the ``$\beta=0$'' row,
are the values of $\lambda_3, \lambda_0$ from simulations of
pure three-dimensional quantum gravity on the same-sized triangulations
\cite{pure3d}.
In these pure quantum gravity simulations a phase transition was found at
$\lambda_0 \approx 0.61$;
for $\lambda_0 < 0.61$ we are in a ``strong gravity'' phase (Newton's constant
$G \sim {1 \over \lambda_0}$ is large) where the fractal or Hausdorff dimension
$d_H$
of the triangulation becomes large. (Note $d_H$ should not be confused with
$d$, the fractal dimension
of the Ising spin clusters.)
In Fig. 6 we see a roughly linear relationship between the $\lambda_3,
\lambda_0$ values and $\beta$
with no dramatic change at the Ising model phase transition around
$\beta_c=0.4$.
The Ising model reduces the values of both $\lambda_3, \lambda_0$ and therefore
has the effect of reducing the cosmological constant and
making already strong gravity stronger.
Perhaps this makes sense - matter does couple positively to gravity after all,
and the cosmological constant is essentially zero in the real universe.
However, it is also apparent that there are large finite-size corrections
which tend to increase $\lambda_3$ and decrease $\lambda_0$.
Therefore when large simulations are done it may turn out that
$\lambda_0$ remains at its pure quantum gravity value despite the addition
of matter in the form of Ising spins, whereas $\lambda_3$ is significantly
reduced, perhaps to zero.

\begin{center}
\begin{tabular}{|c|c|c|c|c|} \hline
$\beta$ & $\lambda_3 (393)$ & $\lambda_0 (393)$ & $\lambda_3 (878)$ &
$\lambda_0 (878)$ \\[.05in]
\hline
 0      & 1.38            & 0.24            & 1.40              & 0.36
\\[.05in]
 0.10   & 1.25(1)         & 0.19(1)         & 1.202(8)          & 0.360(8)
\\[.05in]
 0.20   & 1.16(1)         & 0.15(1)         & 1.101(8)          & 0.337(8)
\\[.05in]
 0.30   & 1.09(1)         & 0.12(1)         & 1.017(8)          & 0.319(8)
\\[.05in]
 0.35   & 1.04(1)         & 0.11(1)         & 0.961(8)          & 0.314(9)
\\[.05in]
 0.40   & 1.03(1)         & 0.10(1)         & 0.945(8)          & 0.308(9)
\\[.05in]
 0.45   & 0.97(1)         & 0.06(1)         & 0.900(8)          & 0.284(8)
\\[.05in]
 0.50   & 0.96(1)         & 0.05(1)         & 0.875(8)          & 0.253(8)
\\[.05in]
\hline
\end{tabular}
\end{center}
{\narrower\smallskip\noindent
Table 5: Predicted values of $\lambda_3, \lambda_0$ from grand canonical dual
simulations
with $N_3 = 393$ and $878$.
\smallskip}
\bigskip

In addition to $E,M,a_M,a_f$ we also measured the average curvature of the
triangulation. In the strong gravity regime for pure quantum gravity
this is found to be negative
\cite{pure3d} and indeed we find this here.
The values of $E,M$ and $a_M$ for the grand canonical simulation are
consistent with those for the micro-canonical;
therefore, despite doubts about its ergodicity, the micro-canonical simulation
does provide a good approximation to the grand canonical.
The values of $a_f$ are substantially reduced because now this quantity
includes the acceptance rate for the
Alexander moves as well as the flip moves, and the former are greatly
suppressed by the relatively large change in energy brought about by
the creation and destruction of (in this case three) Ising spins.

The fractal dimension of the clusters constructed by the Wolff algorithm,
remains close to that found for the micro-canonical simulation, namely a little
less that the value for the standard 3D Ising model. This is probably due
to the fact that the underlying triangulation has a large Hausdorff dimension,
therefore clusters of spins built upon it grow more slowly than those
built with an underlying 3D lattice.

\begin{center}
\begin{tabular}{|c|c|c|c|c|c|} \hline
$\beta$ & $E$     & $M$      & $a_M$    & $a_f$     & $\bar R $ \\[.05in]
\hline
 0.10   & 0.903(1) & 0.034(1) & 0.85(3) & 0.107(12) & -0.333(16) \\[.05in]
 0.20   & 0.784(1) & 0.045(1) & 0.69(3) & 0.105(12) & -0.383(16) \\[.05in]
 0.30   & 0.655(1) & 0.068(1) & 0.54(3) & 0.097(12) & -0.409(17) \\[.05in]
 0.35   & 0.585(1) & 0.099(2) & 0.47(4) & 0.087(11) & -0.414(17) \\[.05in]
 0.40   & 0.494(1) & 0.221(4) & 0.38(5) & 0.077(12) & -0.411(18) \\[.05in]
 0.45   & 0.270(1) & 0.713(1) & 0.20(5) & 0.054(11) & -0.448(17) \\[.05in]
 0.50   & 0.147(1) & 0.871(1) & 0.10(3) & 0.041(8)  & -0.415(17) \\[.05in]
\hline
\end{tabular}
\end{center}
{\narrower\smallskip\noindent
Table 6: Measured values of energy $E$, magnetization $M$,
Metropolis acceptance $a_M$, flip acceptance $a_f$ and
average curvature $\bar R$ from $N_3=878$ grand
canonical dual simulation.
\smallskip}
\bigskip

We have not yet investigated the more physically relevant
case of ``weak gravity'' where Newton's constant is small
and $\lambda_0$ large (at least $>0.61$).
This would involve fixing $\lambda_0$ and hence $N_0$, then performing a
canonical simulation in which $\lambda_3$ and $N_3$ are allowed to vary.
Here we have instead let the system pick its preferred value of $\lambda_0$.

\section{Conclusions}

We have investigated the properties
of the Ising model coupled to three-dimensional quantum gravity.
Firstly, from micro-canonical simulations, we have determined that
the dual implementation of Ising spins in the tetrahedra of the triangulation
is more effective than the direct implementation on vertices, and that
both cases exhibit a phase transition which is most likely second order.
Then, for the dual implementation, we performed a grand canonical
simulation which also displays a second order phase transition,
and discovered that the effect of the Ising spins is to definitely
reduce the cosmological constant and perhaps make already strong
quantum gravity stronger.

The next step in these simulations is to exhaustively explore the parameter
space by choosing fixed values for $\lambda_0$ both in the weak and strong
pure quantum gravity phases, and -- of course -- to simulate larger systems
so that finite-size scaling analysis can be done to
obtain critical exponents and extrapolate to the continuum limit.

\bigskip
\bigskip
\bigskip
\begin{center}
{\bf Acknowledgements}
\end{center}
\bigskip
CFB was supported by DOE under contract DE-AC02-86ER40253
and by AFOSR Grant AFOSR-89-0422.
I would like to thank M. E. Agishtein for help in the initial
stages of this work and A. A. Migdal for useful discussions.

\eject

\begin{center}
{\bf Figure Captions}
\end{center}
\bigskip
\begin{description}
\item{Fig. 1.}
Specific heat for direct micro-canonical simulation.
\item{Fig. 2.}
Susceptibility for direct micro-canonical simulation.
\item{Fig. 3.}
Specific heat for dual micro-canonical and grand canonical simulations.
\item{Fig. 4.}
Susceptibility for dual micro-canonical and grand canonical simulations.
\item{Fig. 5.}
Cluster size vs. diameter, at closest $\beta$ values to $\beta_c$,
for direct $N_0=325$ and dual $N_3=878$ micro-canonical simulations,
along with fits to $d$ shown as dotted and full lines respectively.
\item{Fig. 6.}
Predictions for $\lambda_0,\lambda_3$ from grand canonical simulations.
\end{description}

\eject

\end{document}